\def\shs{2MASS~J09440940-5617117\ }
\def\sshs{2MASS~J09440940-5617117}

\documentstyle[twoside, epsf]{article}

\input{ibvs2.sty}

\begin{document}
\IBVShead{}{}

\IBVStitle{Differential photometry of \sshs\footnote{Based on observations made at the Observat\'orio do Pico dos Dias,
Brazil, operated by the Laborat\'orio Nacional de Astrof\'\i sica.}}

\IBVSauth{K. M. G. Silva$^1$, C. V. Rodrigues$^1$, F. J. Jablonski$^1$, F. D'Amico$^1$, D. Cieslinski$^1$, R. Baptista$^2$ and L. A. de Almeida$^1$}

\IBVSinst{Instituto Nacional de Pesquisas Espaciais, Divis\~ao de Astrof\'\i sica, Brazil}
\IBVSinst{Universidade Federal de Santa Catarina, Departamento de F\'\i sica, Brazil}

\SIMBADobjAlias{\sshs}{[PK2008]HalphaJ094409}
\IBVStyp{Cataclysmic variable}
\IBVSkey{photometry}
\IBVSabs{\shs was identified as a cataclysmic variable in 2007 by its spectrum.}
\IBVSabs{In this article we present optical differential photometry in R$_C$ band obtained using the 0.6-m telescope at the Observat\'orio do Pico dos Dias/Brazil.}
\IBVSabs{The complete light curves confirm the presence of a deep V-shaped eclipse. We derive an orbital period of 0.1879340(5) d. The eclipse has a width of 0.112 $\pm$ 0.003 orbital cycles.}
\begintext

Cataclysmic variables (CVs) are binary stars consisting of a white dwarf accreting matter from a low mass companion via Roche-lobe overflow. \shs was identified as a cataclysmic variable by Pretorius \& Knigge (2008) using the SuperCOSMOS H$\alpha$ survey (Parker et al. 2005). They performed time-resolved spectroscopy from which a probable orbital period of 0.1877(2) d was estimated. The spectrum shows emission lines of the Balmer series and Helium with strong HeII$\lambda$ 4686. They have also presented photometry, which does not cover the entire orbital period. Both photometry and spectroscopy indicate an eclipsing system. By the observational characteristics of this system, Pretorius \& Knigge (2008) tentatively suggest a SW Sex classification. 

We obtained optical photometry of \shs in 2008 at Observat\'orio do Pico dos Dias (OPD) operated by the Laborat\'orio Nacional de Astrof\'\i sica in Brazil. The data were obtained with the 0.6-m Boller \& Chivens telescope at OPD in three nights. The CCD arrays used are 1024 x 1024 pixels back-illuminated SITe devices. Table \ref{parametro2} presents a log of the observations. Figure 1 shows the observed field-of-view around \shs.    

We have used IRAF to correct for bias and flat-field and to perform differential photometry. To illustrate the photometric quality, we present in Figure 2 the light curve obtained on March 03, 2008 for \shs and for a comparison star. In this light curve we see differences in the egress of eclipses. The reference star used is USNOB 8593-02515-1, for which the R$_C$ magnitude was estimated in 11.55 $\pm$ 0.15, based on the USNO magnitudes of 593 stars in the same field-of-view.

The data set contains four eclipses, allowing us to determine an ephemeris for the eclipses in the system. We have included the data from Pretorius \& Knigge (2008) to improve the orbital period estimate. Three different methods were used to estimate the period: Phase Dispersion Minimization, String-Length and Discrete Fourier Transform. The best ephemeris for the times of mid-eclipse is: 
\begin{equation}
T_{mid-eclipse} (HJD) = 2\,454\,516.703\,9\;(3) + 0.187\,934\,0\;(5)\;E\; .
\label{eq-ephem}
\end{equation}
The uncertainty in the period was obtained from the spread of the values given by the three different methods. It is a conservative value since this error is twice as large as the one estimated using the expression of Gilliland \& Fisher (1985) considering the noise. Our period estimate is consistent with the previous suggestion of Pretorius \& Knigge (2008). Figure 3 shows the photometric data plotted in phase with our ephemeris. 

The eclipse width ($\Delta \phi$) of \shs is 0.112 $\pm$ 0.003 orbital cycles. It was calculated considering the phases of minimum and maximum derivative of the mean light curve, indicated by the dotted lines in Figure 4. 

In a CV with a geometrically thin disk, the eclipse width and the mass ratio (q=$M_2/M_1$) can be used to estimate the inclination (i) of the system, as shown by Horne (1985). From the orbital period, we have obtained an estimate of the mass of the secondary star (see Table 2) using the table presented by Knigge (2006). Considering a wide range of white dwarf masses, 0.35-0.77 $M_{\odot}$, we have constructed a diagram of orbital inclination versus mass ratio, which is shown in Figure 5. For the estimated eclipse width, the lower limit to the mass ratio of the system is 0.66, while the upper limit can be found considering the limit of stable mass transfer (q $\textless$ 5/6) and corresponds to 0.83. Considering these limits, the orbital inclination range is 84-90$^{\circ}$. We remark that these results rely on the assumption that the disk is geometrically thin and that its center of light coincides with the white dwarf. This assumption fails if the accretion disk of \shs is geometrically thick and suffers self-occultation
-- as it seems to occur in some SW Sex stars (Knigge et al. 2000).

\textit{Acknowledgements}. We acknowledge the referee, C. Knigge, for his comments. C.V. Rodrigues and K.~M.~G. Silva acknowledge CNPq and FAPESP grants, Procs. 308005/2009−0 and 2008/09619−5, respectively.

\references

Gilliland R. L., Fisher R., 1985, PASP 97, 285 \BIBCODE{1985PASP...97..285G}

Horne K., 1985, MNRAS 213, 129 \BIBCODE{1985MNRAS.213..129H}

Knigge C. , 2006, MNRAS, 373, 484 \BIBCODE{2006MNRAS.373..484K}

Knigge C. et al., 2000, ApJ, 539, 49 \BIBCODE{2000ApJ...539L..49K}

Parker Q. A.  et al.,  2005, MNRAS, 362, 689  \BIBCODE{2005MNRAS.362..689P}

Pretorius M. L., Knigge C., 2008, MNRAS, 385, 1471 \BIBCODE{2008MNRAS.385.1471P}

\endreferences

\IBVSfig{5cm}{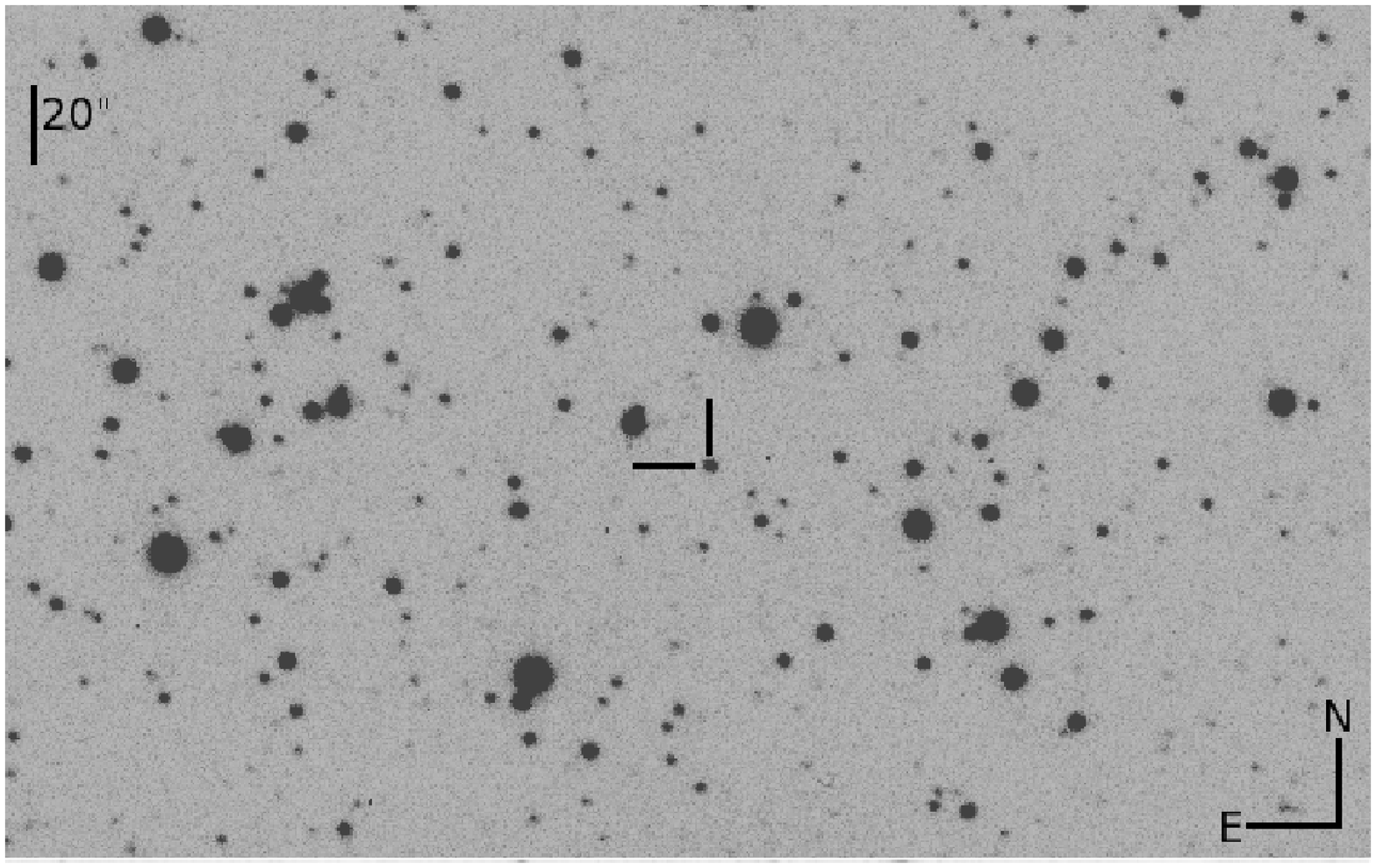}{Finding chart for \shs in the R$_c$ band.}
\IBVSfig{8cm}{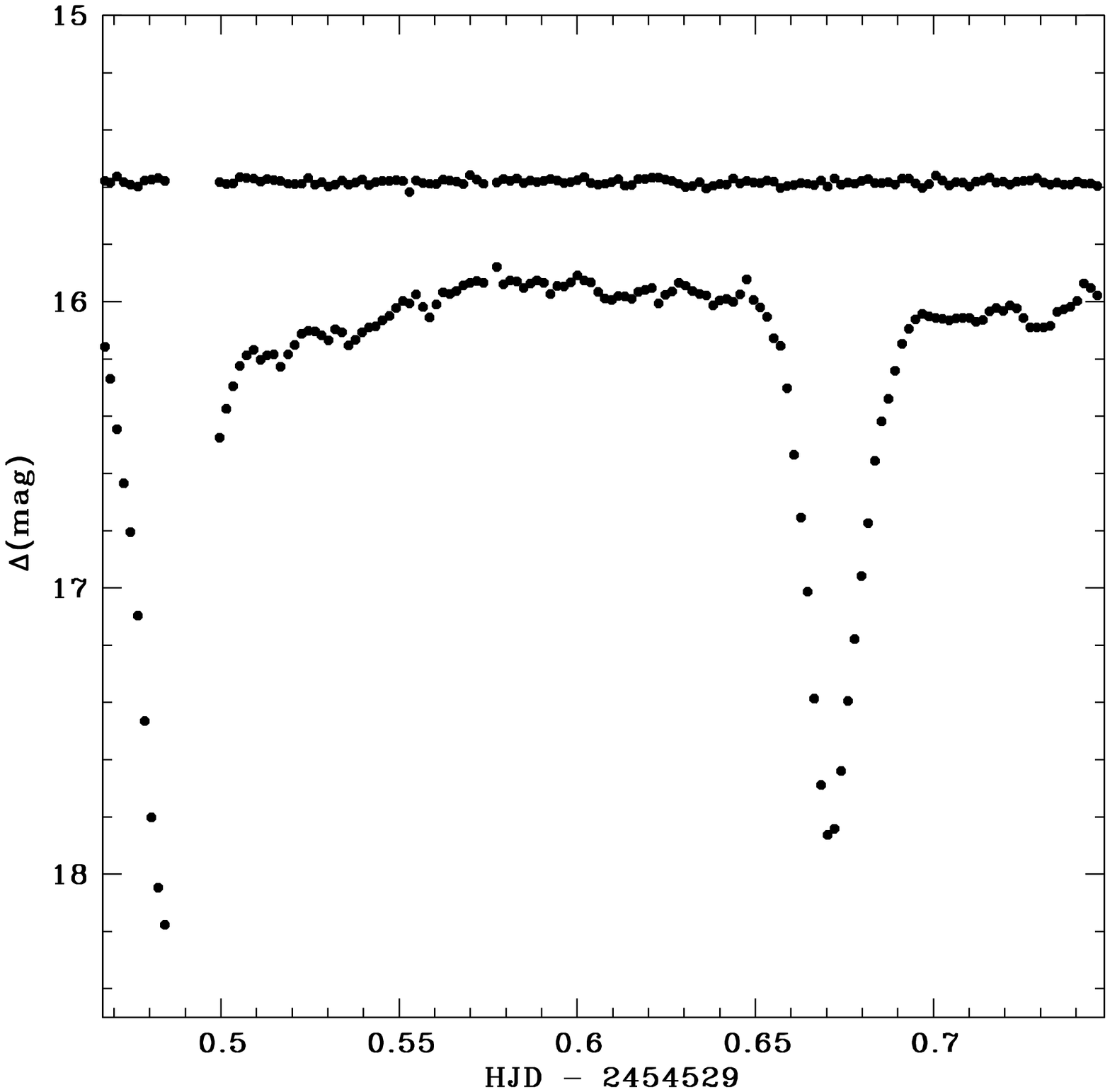}{Optical light curve in the R$_c$ band of \shs on March 03, 2008. The light curve of a comparison star is also presented.}
\IBVSfig{8cm}{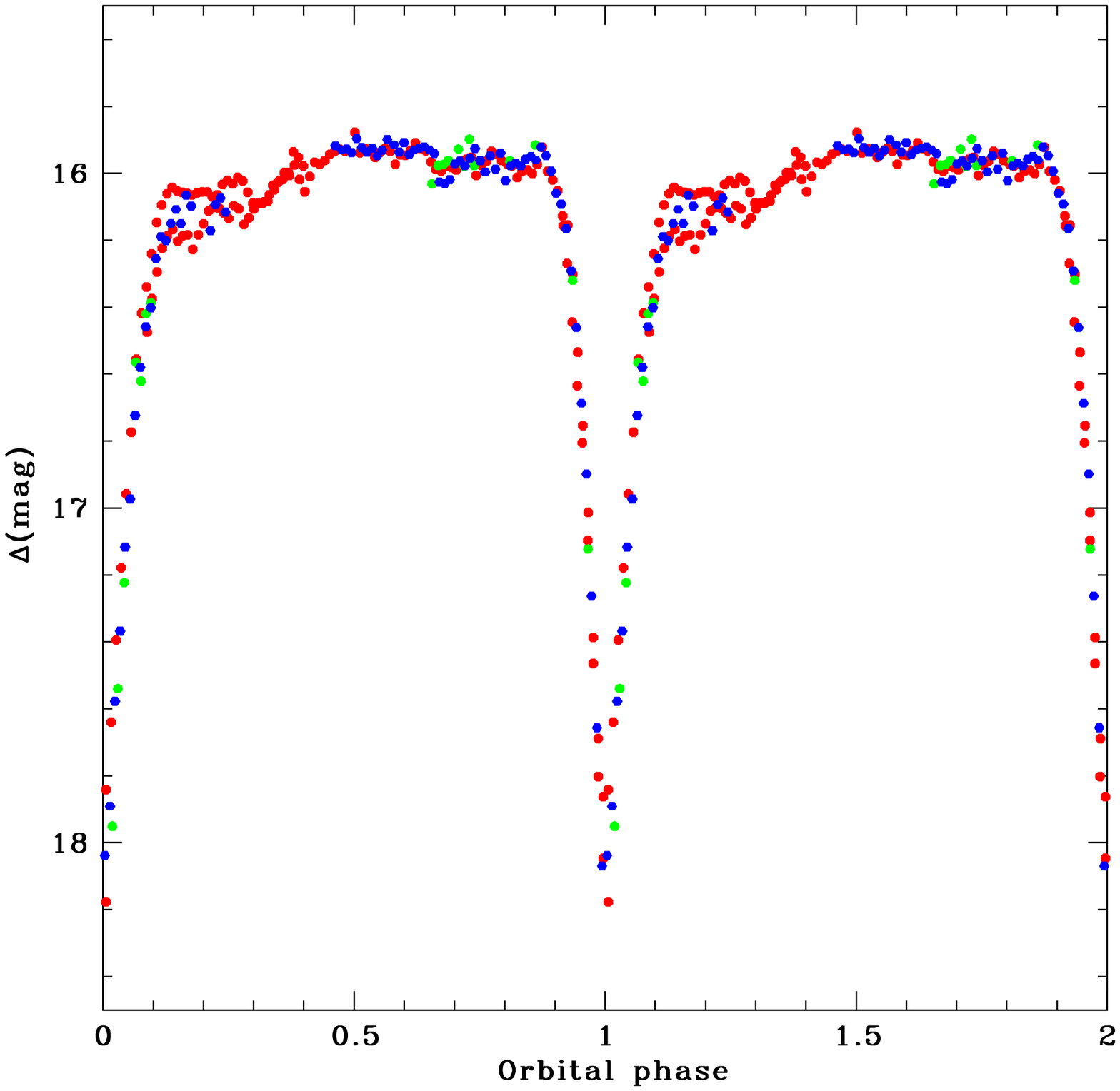}{Phase diagram of \shs in the R$_c$ band on March 03, 2008 (red), on February 19, 2008 (blue) and on February 18, 2008 (green).}
\IBVSfig{7.0cm}{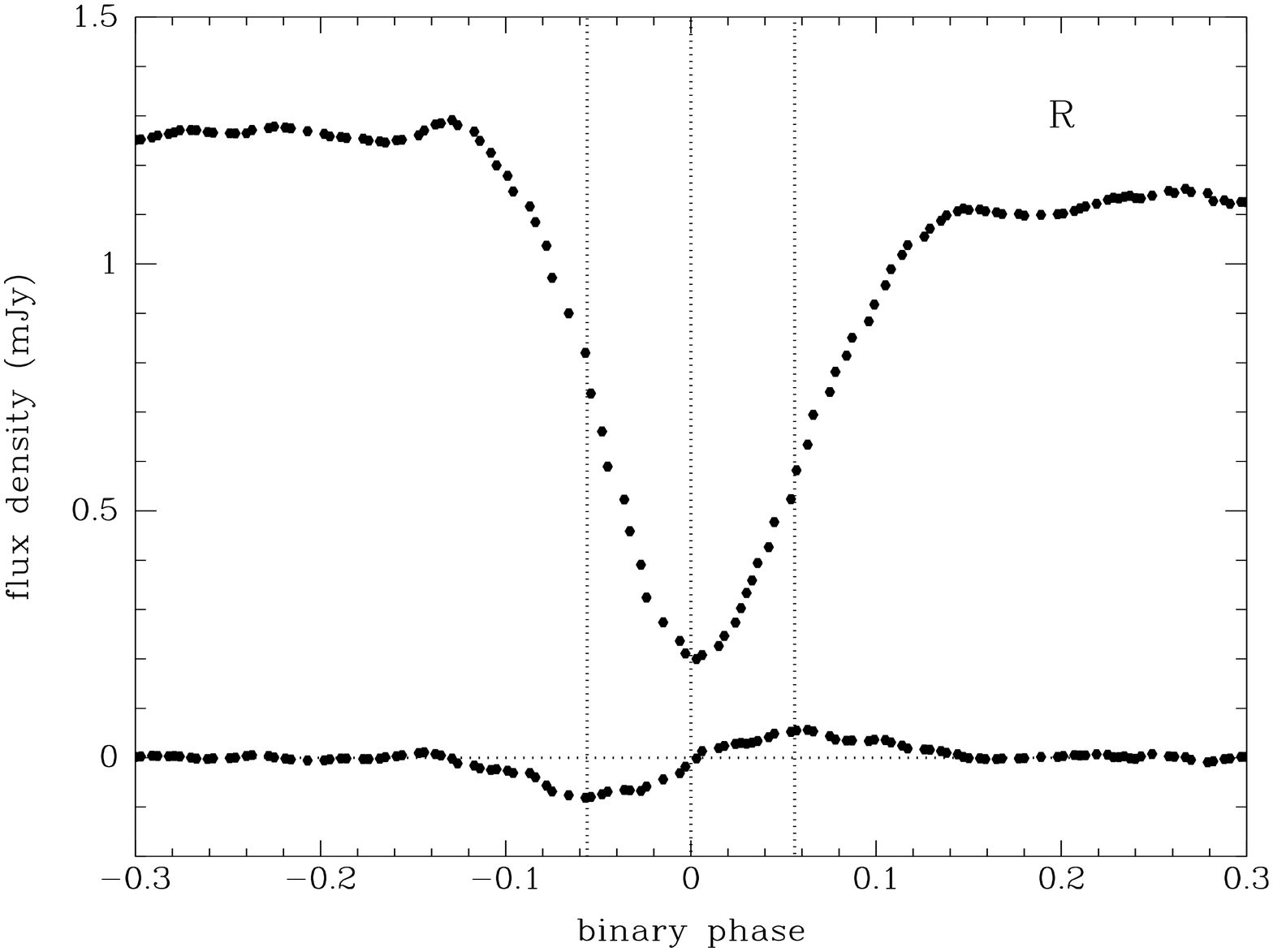}{Mean eclipse profile and its derivative. The dotted lines indicate the center of the eclipse and the phases of minimum and maximum of the derivative.}
\IBVSfig{7.0cm}{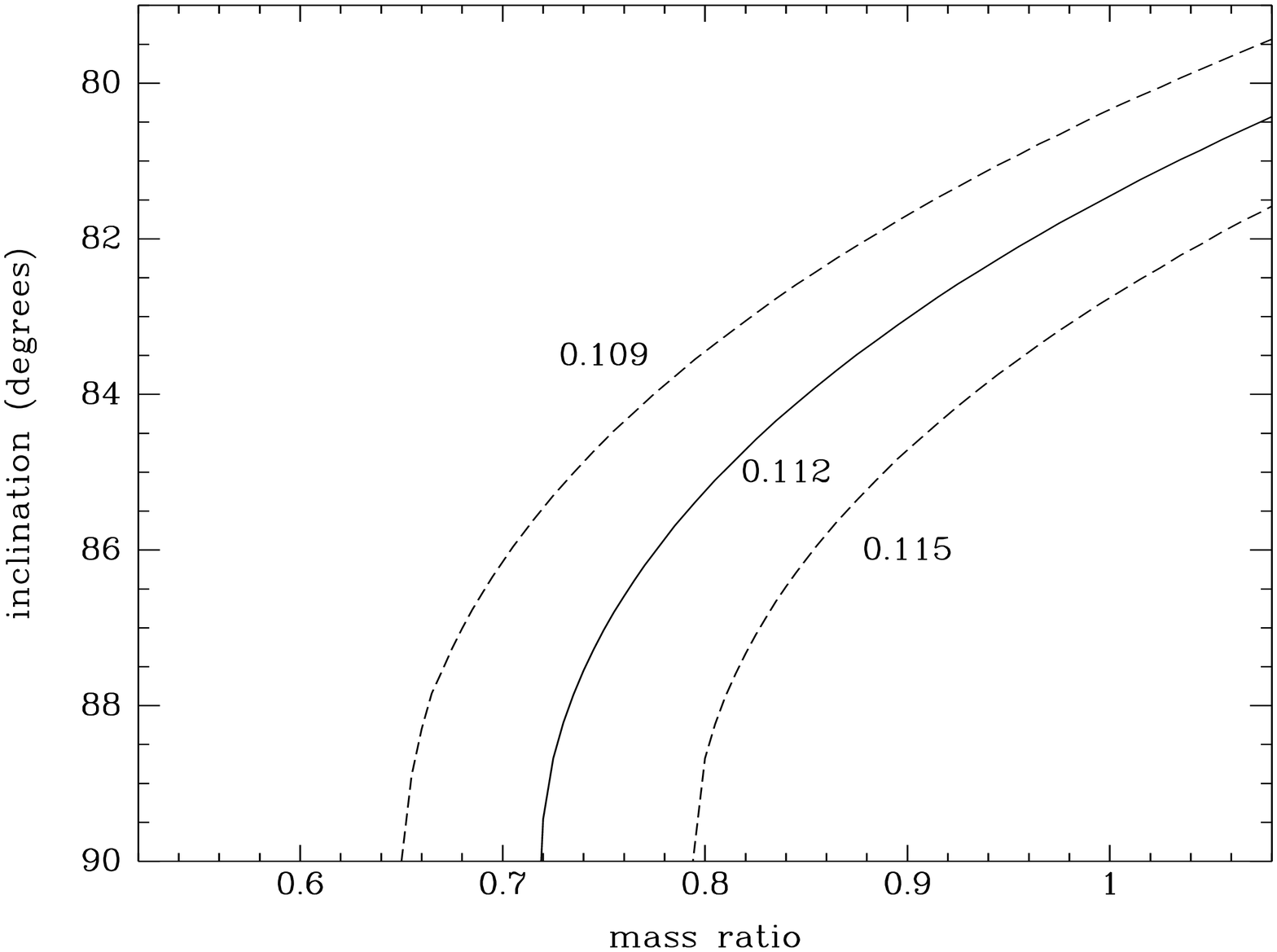}{Orbital inclination versus mass ratio for an eclipse width of 0.112 $\pm$ 0.003 orbital cycles.}

\begin{table}
\begin{center}
\caption{Log of observations}
\label{parametro2}
\begin{tabular}{l c c c c }
\hline
Date & Telescope & Filter & Exposure time (s) & Number of images \\
\hline
2008 Feb 18 & OPD/0.6m & R$_C$ & 120 & 19\\
2008 Feb 19 & OPD/0.6m & R$_C$ & 120 & 74 \\
2008 Mar 03 & OPD/0.6m & R$_C$ & 120 & 140 \\
\hline
\end{tabular}
\end{center}
\end{table} 

\begin{table}
\begin{center}
\caption{Parameters of \shs}
\label{parametro}
\begin{tabular}{l r l }
\hline
Parameter &  & Comments\\
\hline
$P_{orb}$ & 0.1879340(5) d & this work \\
$\Delta \phi$ & 0.112 $\pm$ 0.003 & this work\\
q & 0.66-0.83 & this work\\
i & 84-90$^{\circ}$ & this work \\
$M_2$ & 0.4 $M_{\odot}$ & donor sequence - Knigge (2006) \\
\hline
\end{tabular}
\end{center}
\end{table}

\end{document}